\documentclass[%
prl%
 ,twocolumn%
 ,secnumarabic%
,amssymb, amsmath,nobibnotes, aps, prl]{revtex4}
\usepackage{bm}%
\usepackage{epsfig}%
\usepackage{graphicx}%
\expandafter\ifx\csname package@font\endcsname\relax\else
 \expandafter\expandafter
 \expandafter\usepackage
 \expandafter\expandafter
 \expandafter{\csname package@font\endcsname}%
\fi

\begin{document}

\title{Gravitational Leptogenesis}%

\author{Gaetano Lambiase$^{a,b}$,  Subhendra  Mohanty$^{a,b,c}$ }%
\address{$^a$Department of Physics, University of Salerno, 84081 -
Baronissi, Italy.}
\address{$^b$INFN - Gruppo Collegato di Salerno, Italy.}
\address{$^c$Physical Research Laboratory, Ahmedabad 380009,
India.}
\def\be{\begin{equation}}
\def\ee{\end{equation}}
\def\al{\alpha}
\def\bea{\begin{eqnarray}}
\def\eea{\end{eqnarray}}

\begin{abstract}
We introduce a dimension five CP violating coupling between the
Ricci scalar and fermions. This operator splits the energy level
between neutrinos and anti-neutrinos and can generate a
lepton-asymmetry in the the radiation era  if heavy Majorana
neutrinos decouple at the GUT scale. This operator can also
generate a lepton asymmetry during warm inflation if the light
neutrinos have a Majorana mass $m_\nu \simeq 0.25 eV$ which is
observable in double beta decay experiments.
\end{abstract}

\maketitle

The successful prediction of the light element abundances by BBN
\cite{Copi} and the observations of CMB anisotropies \cite{wmap}
show that the baryon to photon number of the universe is $n_B/s=(
9.2 \pm 0.5)\times 10^{-11}$. Explaining this number in terms of
known parameters is one of the major problems in cosmology.

Recently it was pointed out \cite{Davoudias} that baryogenesis can
be achieved by gravitational violation of the baryon number by a
dimension six operator ${\cal L}= (1/M^2)\, \sqrt{-g} \, R\,
\partial_\mu J^\mu$, where $J_\mu$ can be a linear combination of
baryonic and leptonic currents and $R$ is the Ricci scalar.
Sakharov condition \cite{Sakharov} of $CP$ violation is met as
this operator is odd under $CP$. However when $R$ takes a non-zero
value $CPT$ is broken, so Sakharov's condition baryon number
violating interactions being of out of thermal equilibrium is not
required. Due to $CPT$ violation there is an effective chemical
potential which generates baryon-asymmetry at thermal equilibrium.

In this paper we show that the lowest order $CP$ violating
interaction between fermions and background gravity is the
dimension five operator ${\cal L}_{\diagup{\!\!\!\!\!\!C\!\!P} }
=(1/M_P)\,\sqrt{-g}\, R \bar \psi \,i\gamma_5 \psi$, where $M_P=
(8 \pi G)^{1/2}=2.4 \times 10^{18}GeV$ is the reduced Planck mass.
If $\psi$ is a Majorana neutrino, this operator splits the energy
levels of the particles and anti-particles, and in the presence of
lepton number violating interactions, there is a net lepton
asymmetry generated at thermal equilibrium. In a radiation
dominated era, the lepton asymmetry is given in terms of the
decoupling temperature $T_D$ and the equation of state parameter
$\omega=p/\rho$
 \be
 \frac{n_L}{s} \simeq (1-3 \omega)\, \frac{T^4_D}{M^4_P}\,.
 \ee
With $(1-3 \omega)\sim 10^{-2}$ (generated by quantum corrections
to the thermodynamic free energy of the plasma \cite{kajantie}),
and decoupling temperature at the GUT scale, $T_D \sim
10^{16}GeV$, we obtain the value $(n_L/s) \simeq 10^{-10}$ for
lepton asymmetry. As first pointed out  in \cite{Fukugita},
electro-weak sphalerons at the temperatures $T \sim 10^3 GeV$
violate $B+L$ and conserve $B-L$ \cite{Kuzmin}. Therefore, if a
lepton asymmetry is generated at an earlier epoch, it gets
converted to baryon asymmetry of the same magnitude by the
electro-weak sphalerons. We show that if we have heavy neutrinos
$N_R$ and the equilibrium between the lepton number violating
interactions, which change the $N_R$ and $N_R^c$ abundances,
frozen out at the decoupling temperature corresponding to GUT
scale, then the subsequent decays of $N_R$ and $N_R^c$ would give
rise to the required lepton asymmetry even if there is no CP
violation in the decay, i.e. the $CP$ violation in the Yukawa
couplings is negligible. This is different from the standard
scenario  where $n(N_R)=n(N_R^c)$ as
demanded by $CPT$, but $\Gamma(N_R) \not = \Gamma(N_R^c)$ due to
the complex phases of the Yukawa coupling matrix, and a net lepton
number arises from the interference terms of the tree-level and
one loop diagrams \cite{Fukugita,luty,Flanz, vissani}.

We also show that in warm-inflation \cite{Berera} the lepton
number asymmetry can arise from standard model fields. The lepton
number violating $\nu_L \leftrightarrow \nu_L^c$ interactions
arise from an effective dimension five operator which gives rise
to a Majorana mass of the light neutrinos. The lepton asymmetry in
this case can be expressed in terms of the inflation parameters
$\epsilon$ and $H$
 \be
 \frac{n_L}{s} \sim 0.1 \frac {\epsilon H^3}{M_P T_l}\,.
 \ee
Taking $H \sim T_l \sim 10^{13} \, GeV$ and the slow roll
$\epsilon= 0.001$ we can fit the CMB observations and obtain the
required value $(n_L/s) \simeq 10^{-10}$ for the lepton-asymmetry.
The decoupling temperature $T_l$ is related to the light neutrino
mass and $T_l =10^{13} GeV$ implies that the standard model
neutrinos have a Majorana mass $m_\nu=0.25 eV$, which can be
tested in double-beta decay experiments \cite{Vogel}.

We start with the generalization of the energy-momentum operator
for fermions which can violate various combinations of the
discrete symmetries $C$, $P$ and $T$ \cite{okun,khare}
\begin{eqnarray}
T_{\mu \nu}&=& \bar\psi(P_f) [ F_1 P_\mu P_\nu+ F_2 \sigma_{\mu
\alpha } q^\alpha P_\nu + F_3 \gamma_5  \sigma_{\mu \alpha }
q^\alpha P_\nu \nonumber \\[8pt]
&+& F_4 (q_\mu q_\nu - g_{\mu \nu})
 +F_5 \gamma_5 (q_\mu q_\nu - g_{\mu \nu}) ] \psi(P_i) \nonumber\\
[8pt]&+& (\mu \leftrightarrow \nu)\,, \label{T}
\end{eqnarray}
where $P=P_f+P_i$ and $q=P_f-P_i$. Lorentz invariance in the
local-inertial frame demands that $q^\mu T_{\mu \nu}=q^\nu T_{\mu
\nu}=0$. The coefficients of the form factors $F_3$ and $F_5$ are
odd under $CP$.

In cosmological applications the fermion interactions take place in
a local inertial frame. The metric in the neighborhood of any point
$x^\mu$ can be expressed as
\begin{equation}
g_{\mu \nu}(y)= \eta_{\mu \nu}(x) + \frac{1}{2} R_{\mu \alpha \nu
\beta}(x)\, (x-y)^{\alpha}(x-y)^{\beta}+ \ldots \,. \label{g}
\end{equation}
The space is locally inertial in the sense that the Christoffel
connections $\Gamma_{\alpha \beta}^{\mu}(x)=0$ and
$\bigtriangledown_\nu= \partial_\nu$. Gravitational interactions
of fermions in a local inertial frame will be given by the
coupling of the form
\begin{equation}
{\cal{L}}_{int} = g^{\mu \nu}T_{\mu \nu}\,.
\end{equation}
We take the stress tensor operator in (\ref{T}) and contract it with
the locally flat metric (\ref{g}). Using $q_\mu x^{\nu}=
\delta^{\nu}_{\mu}$ and taking the limits $y \rightarrow x$ we
obtain the expression for the gravitational interaction  of fermions
in a local inertial frame, allowing for CP violations,
\begin{equation}
{\cal{L}}_{int}= F_4 R(x) \bar \psi(x) \psi(x) + F_5 R(x)\bar
\psi(x) i \gamma_5 \psi(x)\,.
\end{equation}
So in a local inertial frame we could have a CP even gravitational
coupling given by the coefficient of $F_4$ and a CP violating
gravitational interaction given by the coefficient of $F_5$. On
dimensional grounds we can expect
\begin{equation}
F_4 \sim F_5 \equiv \beta \sim \frac{1}{M_P}\,.
\end{equation}
The $F_4$ term adds a small correction to the mass of the fermions
and we will not pursue its implications in this paper. The CP
violating $F_5$ term will has more interesting consequences.

The leading order (in coupling $1/M_P$) $CP$ violating interaction
between fermions and gravity is given by the dimension five
operator
\begin{equation}
{\cal L}_{\diagup{\!\!\!\!\!\!C\!\!P} }= \sqrt{-g}\,\beta \, R
\bar \psi \,i\gamma_5 \psi\,. \label{cpv}
\end{equation}
This operator is invariant under Local Lorentz transformation and
is even under $C$ and odd under $P$ and conserves $CPT$. We shall
assume that the coupling $\beta$ is identical for all fermions,
i.e. gravity is flavor blind. In a non-zero background $R$, there
is an effective $CPT$ violation for the fermions.
Using the Dirac equation for the four component fermion $\psi$
\begin{equation}
i \gamma^\mu \partial_\mu \psi-m \psi -\beta R \bar \psi
\,i\gamma_5 \psi =0\,,
\end{equation}
we obtain the dispersion relation
\begin{equation}
E^2 \psi=({\bf p}^2 +m^2 + \beta^2 R^2)\psi -\beta(\gamma_5
\gamma^\mu \partial_\mu R)\psi\,.
\end{equation}
For a spatially homogenous background, the energy levels of the
left and right handed fermions split as
\begin{equation}
E^2 \psi_{\pm}= ({\bf p}^2 +m^2 + \beta^2 R^2  \mp \beta \dot R
)\psi_\pm\,,
\end{equation}
where $\psi_{\pm}=(1/2)(1\pm\gamma_5)\psi$ and over-dot represents
time derivative. Now consider a Majorana neutrino in the chiral
representation $\nu_M=(\nu, i\sigma_2
\nu^*)^T=(\nu_L,{\nu_L}^c)^T$. $\dot R \neq 0$ implies a
spontaneous CPT violation. The energy levels of the two component
left handed neutrino $\nu_L=\psi_+$ and the right handed
antineutrino $(\nu_L)^c=\psi_{-}$ are
 \be
 E_{\pm}=\sqrt{{\bf p}^2 +m^2 + \beta^2 R^2}\,  \mp \, \frac{\beta \dot R}{ 2 \sqrt{{\bf p}^2
 +m^2 + \beta^2 R^2}}\,.
 \ee
This generates a difference between neutrinos and antineutrinos at
thermal equilibrium, and a lepton number asymmetry given by
\begin{eqnarray}
n_L &=&  n(\psi_+)-n(\psi_-) \nonumber \\
 &=& g_\nu \int \frac{d^3p}{(2 \pi)^3}
 \left[\frac{1}{1+e^{\frac{E_{+}}{T}}}-\frac{1}{1+e^{\frac{E_{-}}{T}}}\right] \nonumber \\
 &=&  \frac{1}{2 \pi^2}g_\nu \,  T \, \beta \dot R \, f(\frac{m}{T})\,,
\end{eqnarray}
with
 \bea
 f\left(\frac{m}{T}\right)&=&\ln(1+e^{m/T})-\frac{e^{m/T}}{e^{m/T}+1}\,\frac{m}{T}\,,
 \label{fmt}
 \eea
where $g_\nu$ is the number of neutrino flavors and $T$ is the
temperature at which the lepton number interactions which convert
$\psi_+\leftrightarrow \psi_-=\psi_+^c $ are taking place at
equilibrium. In our applications for both the heavy right handed
neutrinos as well as the standard model left handed neutrinos,
$m/T \ll1,$ and in that limit $f(m/T)\simeq \ln(2)$. We have also
assumed that $\beta R << m$, otherwise $m$ in equation (\ref{fmt})
should be replaced by $\sqrt{m^2+\beta^2 R^2}$. A comment is in
order: The interaction (\ref{cpv}) also splits the energy levels
of the left and right handed charged fermions, for example the
dispersion relation of $e^-_L$ and $e^-_R$ will be different. This
does not however generate a lepton asymmetry as $e^-_L$ and
$e^-_R$ carry the same lepton number unlike  Majorana neutrinos
$\nu_L$ and $\nu_R=\nu_L^c$. There is a energy level split between
$e^-_L$ and $e^+_R$ but this does not give rise to a net lepton
asymmetry due to charge conservation.

The lepton number $n_L$ to entropy $s(=0.44 g_{*s} T^3)$ ratio
will be frozen at the decoupling temperature $T_D$ when the
interaction rate falls below the Hubble expansion rate
 \be
 \frac{n_L}{s} \simeq \, 0.08 \,\frac {g_\nu}{g_{*s}}\, \frac{\beta \dot R}{T^2} \,
 \Big|_{T_D} \label{eta}\,,
 \ee
where $g_{*s}\sim 106.7$ is the effective number of relativistic
particles at the time of decoupling.

{\it Leptogenesis by CP conserving heavy neutrino decays}. GUT
containing the $SO(10)$ group will contain a singlet right handed
neutrino $N_R$ (which is assigned a lepton number $L(N_R)=-1$)
with a large Majorana mass. The heavy right-handed Majorana
neutrino interactions, relevant for leptogenesis, are described by
the lagrangian
 \be
 {\cal{L}}=- h_{\alpha \beta}(\tilde{\phi^\dagger}~
 \overline{N_{R \alpha}} l_{L \beta})
-\frac{1}{2}N_R^c \,M\,N_R +h.c. \,, \label{LN}
 \ee
where $M$ is the right handed neutrino mass-matrix, $l_{L \alpha}=
(\nu_\alpha , e^-_\alpha)_L^T $ is the left-handed lepton doublet
($\alpha $ denotes the generation), $\phi=(\phi^+,\phi^0)^T $ is
the Higgs doublet. In the scenario of leptogenesis introduced by
Fukugita and Yanagida \cite{Fukugita}, lepton number violation is
achieved by the decays $N_R \rightarrow \phi + l_L$ and also
${N_R}^c \rightarrow \phi^\dagger + {l_L}^c$. The difference in
the production rate of $l_L$ compared to $l_L^c$, which is
necessary for leptogenesis, is achieved via the $CP$ violation. In
the standard scenario, $n(N_R)=n(N_R^c)$ as demanded by $CPT$, but
$\Gamma(N_R \rightarrow l_L + \phi) \not = \Gamma(N_R^c
\rightarrow l_L^c + \phi^\dagger)$ due to the complex phases of
the Yukawa coupling matrix $h_{\alpha \beta}$, and a net lepton
number arises from the interference terms of the tree-level and
one loop diagrams \cite{luty, Flanz, vissani}. In our model the
number of $N_R$ and $N_R^c$ at thermal equilibrium differ,
$n(N_R)\not=n(N_R^c)$ as there is a non-zero background value of
$\dot R$. The heavy neutrino lepton asymmetry
$[n(N_R)-n(N_R^c)]/s$ is equal to the asymmetry of the light
neutrinos $[n(\nu_L)-n(\nu_L^c)]/s$ even when decay widths
$\Gamma(N_R \rightarrow l_L + \phi) = \Gamma(N_R^c \rightarrow
l_L^c + \phi^\dagger)$, i.e. even if $CP$ violation in the decays
arising from complex coupling constants are small. In the
radiation era
 \be
 {\dot R}=- (1-3\omega) \,\frac{\dot \rho}{M_P^2}\,.
 \ee
In a $SU(N)$ gauge theory with $N_f$ fermion flavors, the
effective equation of state is given by \cite{kajantie, Davoudias}
 \be 1-3 \omega=\frac{5
\alpha^2}{6 \pi^2}  \frac{[N_c+(5/4)
N_f]\,[11/3)N_c-(2/3)N_f]}{2+(7/2)\,[N_c N_f/(N_c^2-1)]}
 \ee
The numerical value of $1-3 \omega$ depends the gauge group and
the fermions, and  lies in the range $(0.01-0.1)$
\cite{Davoudias}. We shall take the conservative limit, $1-3
\omega=0.01$.
%
One gets
 \be
  \dot R=\frac{4}{\sqrt{3}}(1-3 \omega)\left(\frac{g_*
 \pi^2}{30} \right)^{3/2}\, \frac{T^6}{M_P^3}=4.82
 \,\frac{T^6}{M_P^3}\,. \label{rdot}
 \ee
Using (\ref{rdot}) to evaluate the lepton asymmetry (\ref{eta}),
it follows
 \be
  \frac{n_L}{s}=0.011\, \frac{T_D^4}{M_P^4}\,.
  \ee
If the decoupling temperature of the interactions that keep the
heavy right handed neutrinos in thermal equilibrium is $T_D \sim
2\times 10^{16} \,GeV$ (the GUT scale), we obtain the required
value for $n_L/s \sim 10^{-10}$.


The light neutrino asymmetry can be erased by the interactions
$\nu_L +\phi_0 \rightarrow \nu_L^c +\phi_0^\dagger$ with the
standard model Higgs. To prevent the erasure of the lepton
asymmetry by Higgs scattering, we must demand that the lightest
heavy neutrino mass be lower than the decoupling temperature of
the light-neutrino Higgs interaction, which is calculated as
follows. The light neutrino masses arise from an effective
dimension five operator \cite{dim5} which is obtained from
(\ref{LN}) by heavy neutrino exchange
\begin{eqnarray}
 {\cal{L}}= (h\, M^{-1} \, h^T)_{\alpha \beta} \,(\overline{ {l_{L
 \alpha}}^c }~ \tilde{\phi^*})(\tilde{\phi^\dagger}~ l_{L \beta}) +
 h.c.
 \label{dim5}
 \end{eqnarray}
In the electroweak era, when the Higgs field in (\ref{dim5})
acquires a $vev$, $\langle\phi^0\rangle =v=174~ GeV$, this
operator gives rise to a Majorana neutrino mass matrix
 \begin{equation}
 m_{\alpha \beta}=4 v^2 \,(h \, M^{-1} \, h^T)_{\alpha \beta}
 \end{equation}
The interaction rate of the lepton number violating scattering
$\nu_{L} +\phi_0 \leftrightarrow \nu_{R} +\phi_0^\dagger$ is given
by
 \be
 \Gamma = \frac{0.122}{16 \pi} \frac{m_\nu^2 \, T^3}{v^4}\,.
 \ee
The decoupling temperature $T_{l}$ when the interaction rate
$\Gamma (T_l)$ falls below the expansion rate $H(T_l)= 1.7
\sqrt{g_*} ~T_l^2/(\sqrt{8 \pi }M_{p})$
turns out to be
 \be
 T_l= 2 \times 10^{14}\, GeV\,  \left(\frac{0.05 eV}{m_{\nu_3}}
 \right)^2\,,
\label{Tl}
 \ee
where $m_{\nu_3}$ is the mass of the heaviest neutrino. A lower
bound on the  mass of the heaviest neutrino is given by
atmospheric neutrino experiments \cite{super-K} $m_\nu^2 >
\Delta_{atm}= 2.5 \,\, 10^{-3} eV^2$, which means that the
decoupling temperature has an upper bound given by $ T_l < 2\times
10^{14} GeV$. WMAP \cite{wmap} puts an upper bound $m_{\nu_3} <0.7
eV$, which  puts the lower bound  $T_l > 1.0 \times 10^{12} eV$.
So, $T_l$ lies somewhere in the range $10^{12}GeV < T_l <
10^{14}GeV$ depending on the light neutrino mass. In order that
the lepton number in the $N_R$ and $N_R^c$ asymmetry not be wiped
out by the $\nu_l+\phi_0 \rightarrow \nu_L^c +\phi_0^\dagger$
interactions, the heavy neutrinos must decay after the temperature
$T_l$, which means that the decay width $\Gamma(T_l)<H(T_l)$. The
decay width of the lightest heavy neutrino is
 \be
 \Gamma_{N_1}=\frac{(h h^\dagger)_{11}}{8 \pi}\,M_1 \sim \frac{1}{32 \pi v^2} \, m_{\nu_1}\,
 M_1^2\,,
 \ee
where $m_{\nu_1}=\sqrt{\Delta_{\odot}}=0.007 eV$ if the neutrino
masses are hierarchical. In order that the lightest heavy neutrino
$N_1$ decay after the universe has cooled below the decoupling
$T_l$, its mass must have an upper bound,
 \be
 M_1 < 0.77 \, T_l \, \frac{\sqrt{\Delta_\odot}}{m_{\nu_1}}\,,
 \ee
where $T_l= 2 \times 10^{12} GeV$ if the neutrino  masses are
hierarchial.

In order for the heavy neutrinos to attain thermal equilibrium,
the re-heating temperature after inflation must be above the GUT
scale. This can be achieved by the mechanism of pre-heating
\cite{Giudice} by introducing a coupling between the right handed
neutrinos and the inflaton. At the end of inflation, the
non-perturbative decay of inflaton oscillations must produce
right-handed neutrinos with temperature in the GUT scale, for our
mechanism to work.

To summarize, a heavy right handed neutrino which decouples at
$10^{16}GeV$  and decays below the temperature  $10^{12} GeV$ will
produce a lepton asymmetry in the light neutrinos of the required
magnitude, $n_L/s \sim 10^{-10}$. The main pre-requisite is that
there must be thermal distribution of right handed neutrinos at
the GUT temperature.

{\it  Leptogenesis in Warm Inflation.} The natural application of
this scenario is the warm inflation models as there is a large
non-zero Ricci curvature from the inflaton potential during
inflation and a large temperature where the lepton number
violating interaction can be at equilibrium. Only the standard
model fields, the light neutrino and the Higgs doublet, are needed
for generating lepton asymmetry. The CP violation is provided by
the operator (\ref{cpv}) where $\psi=(\nu_L,\nu_L^c)$ in the
chiral representation. The lepton number violating interactions
which change the number of $\nu_L$ and $\nu_L^c$ are $\nu_L
+\phi_0 \leftrightarrow \nu_L^c +\phi_0^\dagger$ arising from the
effective dimension five operator (\ref{dim5}). The decoupling
temperature is related to the light neutrino mass as given in
(\ref{Tl}).

The Ricci scalar is related to the Hubble expansion rate during
inflation as $R=-12 H^2$ and its time derivative is related to the
slow roll parameter $\epsilon =-\dot H/H^2$ as $\dot R=24 \epsilon
H^3 $. The lepton asymmetry in warm inflation is given by
 \be
\frac{n_L}{s}=\frac{n(\nu_L)-n(\nu_L^c)}{s}= 1.91
\frac{g_\nu}{g_*} \frac{\epsilon H^3}{M_P\, T_{l}^2}\,,
\label{eta2}
 \ee
where $T_l$ is the light neutrino decoupling temperature
(\ref{Tl}).

The power spectrum of curvature perturbation in thermal inflation
and the spectral index of scalar perturbations are expressed in
terms of $H$ and $\epsilon$ \cite{Moss}
 \be
 {\cal P}_{\cal R}=\left(\frac{\pi}{16}\right)^{1/2}\,
\frac{H^{1/2}\,\Gamma^{1/2}\,T}{|\epsilon|}\,, \quad
 n_s-1=-\frac{27}{4} \frac{H}{\Gamma} \epsilon\,,
 \ee
where $\Gamma$ is the damping parameter in the inflaton equation
of motion and represents the coupling between the inflaton and the
thermal bath.

Choosing the parameters $(H,\Gamma,T,\epsilon)$ as $H\simeq
7\times 10^{12}GeV$, $\Gamma \simeq 1.5\times 10^{12}GeV$, $ T
\simeq 8 \times 10^{12}GeV $ and $\epsilon=-0.001$ we obtain $
{\cal P}_{\cal R}\simeq 2 \times 10^{-9}$ and $n_s=0.968$ which
are consistent with the WMAP observations \cite{wmap} of the
amplitude of the curvature power spectrum $ {\cal P}_{\cal R}=
(2.3 \pm 0.3) \times 10^{-9}$ and the spectral index $n_s=0.951
\pm 0.017$. Substituting these values of $(H,T,\epsilon)$ in
(\ref{eta2}) we obtain the required value for lepton asymmetry,
$n_L/s= 1.19 \times 10^{-10}$. From (\ref{Tl}) we see that $T_l=
10^{13} GeV$ corresponds to the neutrino mass of $m_{\nu_3}=0.15
eV$.

{\it Discussions} In this paper we have shown that the leading
order fermion-gravity $CP$ violating interaction is given by
(\ref{cpv}). We have shown that for the Majorana neutrinos, this
operator has the effect of splitting the degeneracy between the
particles and antiparticles which can give rise to lepton number
violation in some generic cosmological scenarios without
fine-tuning any extra parameter. In standard inflation, the heavy
neutrinos must be reheated to the GUT temperature possibly by the
mechanism of pre-heating \cite{Giudice}. In warm inflation the
required lepton asymmetry is achieved by generic values of
parameters which fit CMB observations, and there is a definite
prediction that the standard model neutrinos must have a Majorana
mass $m_\nu=0.25 eV$ which can be tested in double-beta decay
experiments \cite{Vogel}.

\acknowledgements

 SM would like to thank Department of Physics, University of Salerno and
INFN - Gruppo Collegato di Salerno for their hospitality and Prof. G Scarpetta for  discussions.

\end{document}